\documentstyle[12pt]{article}
\newcommand{\beq}{\begin{equation}}
\newcommand{\eeq}{\end{equation}}
\newcommand{\gsim}{\lower.7ex\hbox{$
\;\stackrel{\textstyle>}{\sim}\;$}}
\newcommand{\lsim}{\lower.7ex\hbox{$
\;\stackrel{\textstyle<}{\sim}\;$}}

\begin{document}
\begin{titlepage}
\renewcommand{\thefootnote}{\fnsymbol{footnote}}

\begin{center} \Large
{\bf Theoretical Physics Institute}\\
{\bf University of Minnesota}
\end{center}
\begin{flushright}
TPI-MINN-95/15-T\\
UMN-TH-1345-95\\
hep-ph/9505289
\end{flushright}
\vspace{.3cm}
\begin{center}
{ \Large
{\bf Recent Progress in the Heavy Quark Theory}}

\vspace{1cm}

Talk at the V International Symposium on Particles,
Strings
and Cosmology

-- PASCOS --

{\em  Johns Hopkins University, Baltimore, March 22 - 25, 1995}

\end{center}
\vspace*{.3cm}
\begin{center} {\Large
M. Shifman} \\
\vspace{0.4cm}
{\it  Theoretical Physics Institute, Univ. of Minnesota,
Minneapolis, MN 55455}\\
\vspace*{.4cm}

{\Large{\bf Abstract}}
\end{center}

\vspace*{.2cm}
Some recent developments in the heavy quark theory are briefly
reviewed. The main emphasis is put on interrelation between
HQET and OPE. The notion of duality and deviations from duality are
discussed in detail.

\end{titlepage}

\section{Instead of Introduction}

QCD is a messy theory in the strong coupling regime and with no
obvious expansion parameter. If we could choose for Nature we
would definitely pick up something more elegant, preferably in less
dimensions, with more symmetry and so on. Alas, the choice is not
ours. This is {\em the} theory of matter, and we have to learn how to
deal with it as it is.  Although there has been no decisive
breakthrough
in QCD in the recent years, a steady progress takes place in different
directions. My topic today is the progress in the heavy quark theory.

The reasons why the heavy quark is the best probe of the strong
interactions (out of a rather scarce supply we have at the moment)
are  well known.  The road which led people to the understanding of
this fact was not straightforward, though. The history of
development of the
pertinent ideas and  formalisms  is briefly summarized in my recent
talk
\cite{ShifCAQCD} and in excellent reviews \cite{HQETrev}, and I will
save  time,
sparing the audience the $n$-th repetition of  (interesting and
important)  topics that have become routine.
Instead,
I will concentrate on some very recent results.

The development of the theory goes in the direction of new
applications on the one hand, and a deeper understanding of
foundations, on the other. Among the former let me mention
a few which might have a significant   impact on phenomenology.
The first  example of this type is the issue of the higher-order
perturbative corrections in the inclusive semileptonic width of $B$
mesons where a noticeable progress has been reported
\cite{alpha,alpha2,alpha3}. Another example  is
a
new calculation \cite{Vbc} of the power corrections
to the $b\rightarrow c$ transition form factor at zero recoil.

On the theoretical side, we are witnessing how the focus is shifting
away from building particular $1/m_Q$ expansions
towards more general aspects. Surprising though it is, it was not fully
realized that HQET \cite{HQET} is merely a special version of Wilson's
operator
product
expansion (OPE) \cite{Wils}. As this fact gains recognition more and
more
elements
of Wilson's OPE are being adapted in the HQET environment leading
to powerful, and sometimes unexpected, consequences.  For instance,
a conspiracy between the infrared renormalons and condensates
is  common knowledge in all problems where the operator product
expansion can be used \cite{muelller}. The issue of the infrared
renormalons in the context of HQET was raised in Refs.
\cite{bigi1,bb}.
It took some effort to reveal peculiarities of HQET and to figure out
how concretely the same phenomenon of conspiracy works for heavy
quark expansions \cite{bigi1,bbz} (see also \cite{manohar}).  The
research did
not stop at this point, however.

As was already mentioned, practical needs of phenomenology
require inclusion of higher-order perturbative corrections. In such
problems as the weak inclusive decays of the heavy flavors complete
calculation
of even ${\cal O}(\alpha_s^2)$ terms is a notoriously difficult task.
People tried playing with different approximate approaches; the so
called `$b$ graph dominance' prescription due to Brodsky,
Lepage and Mackenzie (BLM) \cite{BLM} is one of the most popular.
Originally the BLM hypothesis was engineered as a scale setting
procedure in the ${\cal O}(\alpha_s)$ terms allowing one to minimize
the coefficients of the ${\cal O}(\alpha_s^2)$ terms without their
complete calculation. This idea was married with the heavy quark
theory
\cite{alpha2}.
At the next stage the renormalon studies and the BLM calculations
fuse giving rise to an approximation which is sometimes called
`resummation of the renormalon chains' \cite{nb}. Using an extended
BLM hypothesis one isolates and sums up an infinite
sequence of graphs which possibly gives an estimate of
higher-order effects. The results  obtained so
far  \cite{alpha3,nb}  are only the beginning of the story. The
extended BLM resummation procedure  has not yet been properly
formulated in the context of OPE. Terms of a very high order in the
`renormalon chain' come from large distances and are not calculable
in this way. The large-distance contribution should be isolated and
treated separately, as a condensate.  The separation is readily done
in Euclidean calculations but is technically hard to implement in
the problems with the Minkowski kinematics, like in the inclusive
$B$ decays. Until this is done the resummation of the renormalon
chains is hardly more than an academic exercise and of a limited
practical value.
Still, this is
obviously a new  promising direction  deserving further efforts
that will,
perhaps,  eventually find its place in the OPE-based theory.

First attempts of the instanton calculations in the inclusive heavy
flavor decays have been reported recently \cite{resum}. The
instanton
contribution manifests itself in the coefficient functions, as a
deviation from the so-called practical version of OPE \cite{NSVZ}.
This is the only
known example of this type which, unfortunately, is very poorly
controllable numerically. At the present stage one can speak, rather,
about theoretical aspects of the instanton contribution as a
laboratory or a testing ground for different effects not seen within
the practical version of OPE. The topic is not mature enough to
warrant a detailed discussion at this Symposium -- the existing
publications leave more questions than answers and should be,
rather, viewed as an invitation for further work. I will comment,
though, on some general aspects of the instanton calculations later
on.

The  ideas people play with  now existed in QCD, in this or that form,
for years. The heavy quark theory, being one of a few branches of
QCD  still in active growth, gave a new life to them. Being unable to
dwell on all these developments I have picked up two samples from
the current flow: one problem is applied, the other is more
fundamental.

\section{Deviations from the Heavy Quark Symmetry at Zero Recoil
and Determination of $V_{cb}$}

The discussion below will hopefully show how fruitful  the
intrusions of the OPE-based methods are (familiar from the QCD sum
rules) for heavy quark theory.

Needless to say  determining $V_{cb}$ is one of the most
important current tasks in particle physics. Since isolated quarks do
not exist it is
impossible to measure $V_{cb}$ directly, without heavily exploiting a
theoretical kitchen. Strictly speaking, there are two different
kitchens; one of them deals with
 extracting  $V_{cb}$ from
experimental data through analysis of   the exclusive $B\rightarrow
D^*
l\nu$
decays in the so called small velocity  limit (slow $D^*$'s).
Extrapolation of the amplitude to the point of zero recoil yields
$|V_{cb}|F_{B\rightarrow D^*}$(zero recoil), where
$F_{B\rightarrow D^*}$ is an effective $B\rightarrow D^*$ transition
form factor. In the small velocity (SV) limit this form factor is close
to unity as a
consequence of the heavy quark symmetry
\cite{Nussinov,Voloshin,Isgu}; deviations from unity are quadratic in
the
inverse heavy quark mass, $1/m_{b,c}^2$ \cite{Voloshin,Luke}.  Our
task  is to calculate these deviations.

To  predict
$(F_{B\rightarrow D^*} -1)$ at zero recoil we
derive a sum rule for the transitions $B\rightarrow D^*$ and
$B\rightarrow$ vector
excitations generated by the axial-vector
current, $A_\mu =\bar b\gamma_\mu\gamma_5c$. If the
momentum carried by the lepton pair is denoted by $q$, the zero
recoil
point is achieved if $\vec q =0$. To obtain the sum
rule we consider the $T$ product
\begin{equation}
h_{\mu\nu} = i\int d^4x
{\rm e}^{-iqx}\frac{1}{2M_{B}}
\langle B |T\{ A_\mu^\dagger (x) A_\nu (0)\}|B\rangle
\label{1}
\end{equation}
where the hadronic tensor $h_{\mu\nu}$ can be systematically
expanded
in $\Lambda_{\rm QCD}/m_{b,c}$. For our purposes
it is sufficient to keep the terms quadratic in this parameter
and to consider only one out of five possible kinematical structures,
namely $h_1$, the only structure surviving for
the spatial components of the axial-vector current, see e.g.
\cite{Chay,Koyrakh}.

Next we use the standard technology of the QCD sum rule approach.
Let us define
\begin{equation}
\epsilon = M_B-M_{D^*}- q_0 =\Delta M - q_0 \, .
\label{2}
\end{equation}
If $\epsilon$ is positive we sit right on the cut. The imaginary part
of the amplitude (\ref{1}) is the sum of the form factors squared
(taken at zero recoil). The sum runs over all possible intermediate
states,
$D^*$ and excitations.  We want to know  the first term in the sum,
$|F_{B\rightarrow D^*}|^2$. Alas, the
present-day QCD does not allow us to make
calculations directly in this domain.

On the other hand, if $\epsilon$ is negative we are below the cut, in
the Euclidean domain. Here the amplitude (\ref{1}) can be calculated
as an expansion in $1/m_{b,c}$
provided that
$
|\epsilon |\gg \Lambda_{\rm QCD} .
$
To get a well-defined expansion in $1/m_{b,c}$ we must
simultaneously  assume
that
$
\epsilon\ll m_{b,c}
$.

The non-perturbative corrections we are interested in are due to the
fact that both, the $c$ quark propagator connecting the points
$0$ and $x$ in Eq. (\ref{1}) and the external $b$ quark lines,
are not in the empty space but are, rather, submerged into a
soft-gluon medium, a light cloud of the $B$ meson. Two parameters
characterizing the properties of this soft medium are relevant for our
analysis. A chromomagnetic parameter
\begin{equation}
\mu_G^2 =\frac{1}{2M_B}\langle B|
\bar b\,\frac{i}{2}\sigma_{\mu\nu}G^{\mu\nu}\,b|B\rangle
\approx \frac{-1}{2M_B}\langle B|
\bar b\,\vec\sigma\vec B\,b|B\rangle
\label{mug}
\end{equation}
measures the correlation between the spin of the $b$ quark inside
$B$ and the chromomagnetic field $\vec B$ created by the light
cloud. The
second parameter is $\mu_\pi^2 =(2M_B)^{-1}\langle B|
\bar b\,(i\vec{D})^2 \, b|B\rangle$ measuring
the average spatial momentum squared of the $b$ quark. Both
parameters are proportional to $\Lambda_{\rm QCD}^2$. That's all
we need for the leading non-perturbative term.

If the amplitude (\ref{1}) is considered in the Euclidean domain
far below the cut (i.e. $-\epsilon \gg \Lambda_{\rm QCD}$) the
distance between the points  $0$ and $x$  is short and we can
expand $h_1$ in $\Lambda_{\rm QCD}^2/m_{b,c}^2$. Actually,  the
whole amplitude contains  more information than we need; the
sum rule sought for is obtained by considering the coefficient
in front of $1/\epsilon$ in $h_1$. In this way we arrive at the
following prediction:
$$
F_{B\rightarrow D^*}^2 + \sum_{i=1,2,...}F_{B\rightarrow excit}^2=
$$
\begin{equation}
1 -\frac{1}{3}\frac{\mu_G^2}{m_c^2}
-\frac{\mu_\pi^2-\mu_G^2}{4}\left(
\frac{1}{m_c^2}+\frac{1}{m_b^2}+\frac{2}{3m_cm_b}
\right) ,
\label{SR}
\end{equation}
where the sum on the left-hand side runs over  excited states
with the appropriate quantum numbers, up to excitation energies
$\sim \epsilon$. (In other words, $\epsilon$ plays the role of the
normalization point. Higher excited states are dual to the graphs with
the
perturbative hard gluon in the intermediate state are neglected
together with
the latter).
All form factors in Eq. (\ref{SR}) are taken at the point of zero recoil.

Let us now transfer the contribution of the excited states to the
right-hand side and account for  the fact
\cite{Bigi2,Vbc} that $\mu_\pi^2 >\mu_G^2$.
Then we get a lower bound on the deviation
of $F_{B\rightarrow D^*}$ from unity,
\begin{equation}
\eta_A - F_{B\rightarrow D^*} >\frac{\mu_G^2}{6m_c^2}\, .
\label{bound}
\end{equation}
Here we included the perturbative one-loop correction
\cite{CGP,Voloshin} so that $1\rightarrow\eta_A$,
\begin{equation}
\eta_A=1
+\frac{\alpha_s}{\pi}\left(\frac{m_b+m_c}{m_b-
m_c}\log{\frac{m_b}{m_c}}-
\frac{8}{3}\right)\approx 0.975\, .
\label{pertur}
\end{equation}
Using the known value of $\mu_G^2$ and $m_c =$1.3 GeV  we
conclude that $F_{B\rightarrow D^*}<0.94$.

Including the $\mu_\pi^2-\mu_G^2$ term and the contribution from
the excited states lowers the prediction for $F_{B\rightarrow D^*}$
making  deviation from  unity more pronounced. If $\mu_\pi^2$
is taken from the QCD sum rule calculation \cite{Ball} the estimate
of $F_{B\rightarrow D^*}$ is reduced to 0.92. As far as the excited
states are concerned a rough estimate of the $D\pi$ intermediate
state can be given \cite{Vbc} implying that
\begin{equation}
F_{B\rightarrow D^*}=0.89\pm 0.03\, .
\label{pred}
\end{equation}
The error bars here reflect only the uncertainty in the excited states.
The parameters $\mu_\pi^2$, $\mu_G^2$, $m_c$ and $\eta_A$ have
their own error bars which I can not discuss here due to time/space
limitations. Some relevant numerical estimates can be found in Ref.
\cite{mane}.

The corrections ${\cal O}(1/m_{b,c}^2)$ to the form factors at zero
recoil have been discussed previously \cite{Fal,Manne}
within an HQET expansion.
In this version, instead of the excited state contribution, one
deals with certain non-local correlation functions which are
basically unknown. The advantage of the sum rule approach
presented above seems obvious: here we deal with only local
operators; all non-locality is hidden in the excited state contribution
which (i) has  definite sign; (ii) its magnitude is known to be
relatively small numerically; (iii) it can be relatively reliably
estimated in models.

\section{OPE and Deviations from Duality}

Many  applications of the heavy quark expansions are based, apart
from the expansion {\it per se}, on  duality. All problems associated
with the inclusive heavy flavor decays belong to this class. Needless
to say,   they are important.  As usual in QCD, a
legitimate formulation of the expansion can be given in the Euclidean
domain. The physically measurable quantities (e.g. the lepton
spectrum in the inclusive $B\rightarrow l\nu X_u$ decays) belong to
the Minkowski domain. We want to extract consequences of the
Euclidean expansion in the Minkowski domain.  A bridge between
these two is provided by duality,
a concept which is very frequently used but is very poorly studied.
The notion is not even clearly defined in the literature,
so that different people ascribe to it different meaning.

Now I will dwell on some general features of the operator product
expansion in QCD with the intent  to at least initiate  discussion of
this God-forgotten topic, deviations from duality.  Many of the
arguments presented below are part of  ongoing research
\cite{SUV}.

To establish an appropriate setting and introduce necessary
terminology let me start from a classic example,
the total cross section of $e^+e^-$ annihilation to
hadrons. More specifically,  assume, for simplicity,
that only $u$ and $d$ quarks exist, they are massless, and the
``electromagnetic current" has the form
\begin{equation}
j_\mu = \frac{1}{2}(\bar u\gamma_\mu u - \bar d\gamma_\mu d ).
\end{equation}
 Inclusion of other quarks or the
quark masses is a technical issue which does not affect the essence of
the construction. Also, the particular choice of the current is  not of
importance.

The central object of our analysis is the $T$ product of two currents
\begin{equation}
T_{\mu\nu} = i \int {\rm e}^{iqx} d^4 x T\{ j_\mu (x) j^+_\nu (0) \}
\equiv (q_\mu q_\nu - q^2g_{\mu\nu} )\,\, T.
\end{equation}
For large Euclidean $q^2$ ($q^2<0$)
\begin{equation}
T=\sum C_n(q;\mu ){\cal O}_n(\mu )
\label{tee}
\end{equation}
where the normalization point $\mu$ is indicated explicitly.
(Strictly speaking, one should  deal with $q^2$ derivatives of
Eq. (\ref{tee})). In what follows we will also use the notation $Q^2
\equiv -q^2$. The sum in Eq. (\ref{tee}) runs over all possible
Lorentz and gauge invariant local operators built from the gluon and
quark fields. The operator  of  lowest (zero) dimension is the unit
operator {\bf I},
then comes the gluon condensate $G_{\mu\nu}^2$, of dimension four;
the four-quark condensate represents the example of dimension-six
operators (for further details see Ref. \cite{Shif}).

The coefficient functions $C_n$ (absorbing the short-distance
contribution) contain both,
perturbative and non-perturbative contributions of different types.
It is convenient to start from the unit
operator whose coefficient accounts for the entire perturbation
theory.

Consider a typical Feynman graph for the vacuum expectation value
of $T_{\mu\nu}$ depicted on Fig. 1. Assume that the momentum
flowing through the graph is Euclidean and large,
$Q\rightarrow\infty$. All subgraphs
in this graph are assumed to be renormalized at $Q$. The final result,
being expressed in terms of the running coupling constant
$\alpha_s(Q)$
is finite. In the standard calculation of the diagram  of Fig. 1 all
virtual
momenta saturating the loop integrals scale with $Q$ as the first
power
of $Q$. For large $Q$ we use the perturbative expressions for
the quark and gluon Green functions and one might say, naively, that
we
arrive at the
standard perturbation theory for  the coefficient of the unit
operator,
\begin{equation}
C_I= \sum_l a_l \alpha_s^l ,
\label{11}
\end{equation}
where $a_l$ are numerical coefficients.
As a matter of fact, the above expression
is not correct theoretically. One should not forget the
following:
in doing the loop integrations in $C_I$ we {\em must} discard the
domain
of virtual momenta below $\mu$, by definition of $C(\mu )$.
Subtracting
this domain from the perturbative loop integrals we introduce in
$C_I$
power corrections of the type $(\mu /Q)^n$ by hand. (If it is possible
to
choose $\mu$ sufficiently small these corrections may be
insignificant
numerically and can be omitted; such a situation will be realized in
what is
called a {\em practical version of OPE}, see below). Needless to say
that the parameter $\mu$ is auxiliary -- the final result for
$\langle T\rangle$ must be $\mu$ independent.

At any finite order the perturbative contribution, eq. (\ref{11}),
is well-defined. At the same time, if the number of loops becomes of
order
$1/\alpha_s$ and each loop carries roughly speaking one and
the same momentum the total momentum $Q$ is shared between
many lines.  This might be called a semi-hard contribution. One can
visualize it in terms of, say, `direct
instantons' \cite{Novi1}. Integration over the sizes $\rho$ of the
direct
instantons is implied; the integral is saturated at $\rho\sim Q^{-1}$.
It should be emphasized that the correspondence between the
Feynman graphs with $1/\alpha_s$ lines and the semi-hard
contributions described above is qualitative. Still, if we rely on this
correspondence we may say that
the characteristic loop momentum scales like $Q/\ln Q$.
It is quite clear that the contribution of this type has the form
\begin{equation}
\Delta C_I \sim \exp {(-C/\alpha_s (Q))} \sim \left(
\frac{\Lambda_{QCD}}{Q}\right)^\gamma
\label{12}
\end{equation}
where $C$ is some positive constant and the exponent $\gamma$
need not be integer. Typically, the numerical value of $\gamma$ is
large. In
this way we get a non-perturbative
contribution to the
coefficient function $C_I$, a non-condensate non-perturbative term.
Such terms may or may not be important numerically depending on
a particular value of $Q$ under consideration. (They are disregarded
in the practical version of OPE which, thus, applies to the values
of $Q$ above a critical point).

This is not the only source of the non-perturbative terms, however.
Consider a situation in which one, or a few lines in the diagram of Fig.
1 are in a special regime -- their momenta are soft in the sense
that the characteristic off-shellness does not scale with $Q$.  These
lines are marked by crosses on Fig. 2.  Graphically one can cut these
lines and treat the product  as a  local operator. In this way we
arrive to the sub-leading operators ${\cal O}_n,\,\, n\neq 0,$
in the expansion (\ref{tee}). For instance, cutting the gluon line we
get $G_{\mu\nu}^2$. The remaining part of the graph
where the loop momenta scale as $Q$ will yield the coefficient
function in front of the corresponding operator ${\cal O}_n$ in the
form of a
series  in $\alpha_s(Q)$. Taking the matrix element
of the sub-leading operator produces the  {\em condensate}
non-perturbative correction.  (Analogously to $C_I$, the
coefficients $C_n$   also receive, generally speaking, a
semi-hard non-perturbative contribution of the type presented in
eq. (\ref{12}). By the same token, in calculating the perturbative part
of
$C_n$ one must discard the domain of virtual momenta $k<\mu$.).

Finally, it is conceivable that there are no hard or semi-hard
lines at all. The external momentum $Q$ is transferred from
the initial to the final vertex through a large number of quanta
(growing like a power of $Q$) and none of the quanta carries
momentum scaling with $Q$ (Fig. 3). Of course, in this situation one
can not
speak of individual quanta, one should rather use the language of
typical field fluctuations transferring the momentum $Q$ without
having any Fourier components with frequencies of order $Q$.
This type of contribution is not seen in OPE truncated at any
finite order. It is reflected in the rate of  divergence of high orders
in the power series.
It is intuitively clear that the corresponding part of the correlation
function  must be
exponentially small, $\sim\exp (-Q)$. A transparent example is again
provided by instantons. This time one has to fix the
size of the instanton $\rho$ by hand,  $\rho\sim\rho_0$. Then this
contribution is ${\cal O}(\exp (-Q\rho_0
))$.
 The relation between exp$(-Q\rho_0)$ piece and the high-order
terms of the
 power (condensate) series is akin  the connection between
$\exp (-1/\alpha_s)$ terms and $l\sim 1/\alpha_s$ orders
in the perturbative expansion.
Indeed, proceeding to  higher-dimension operators in OPE we force
more and
more lines to
become soft, and the limit of only soft lines is correlated with
the operators built from a large number of the gluon and quark
fields.

I hasten to make the following reservation.
It may happen (and actually happens) that some of the loop
integrations are not saturated at $Q$ or $\mu$ but are rather
logarithmic. Logarithms $\ln (Q/\mu )$ occurring in this way
are associated either with the running coupling constant $\alpha_s
(Q)$ or with the anomalous dimensions of the operators at hand.
They will play an important role in the analysis following below.

Summarizing, Wilson's OPE (\ref{tee}) yields $\langle T\rangle$
in
the deep Euclidean domain as an expansion in different parameters.
Purely logarithmic terms
$(\ln Q)^{-l}$ are due to ordinary perturbation theory.  Terms of the
type $(Q^2)^{-k} (\ln Q)^{-\gamma}$ reflect higher-dimension
operators and semi-hard contributions (`direct instantons'). In the
former case the values of $k$ are  integer,  the latter case
may produce non-integer values of $k$.
In the practical version of OPE we neglect the semi-hard
contributions
calculating the coefficient functions perturbatively. All
non-perturbative terms come from condensates within this
approximation.

The perturbative calculation of the coefficient function yields a series
in $\alpha_s$. Everybody knows that the series is factorially
divergent.  First, the number of graphs grows factorially with
the number of loops. This factorial behavior can be correlated with
the instanton-antiintstanton contribution \cite{HOIA}. Second, there
exist
specific
graphs which are factorially large by themselves, the
renormalons \cite{renorm}. As I have already mentioned the latter
received much attention lately,
in connection with various applications of HQET. In the framework of
the operator product expansion the infrared renormalons {\em per
se} play no
role  since they are completely hidden in the condensate
corrections. By construction, in calculating the coefficient functions
we cut off the infrared contribution at $\mu$, never come close to
the Landau pole and, hence, the coefficient functions, calculated
properly, are free from the infrared renormalons. (The ultraviolet
renormalons are not absorbed in the condensate corrections, but they
are harmless anyway.)

Passing to the next level, from the perturbative series to the
condensate power series, we face the same problem since this power
series is also asymptotic and   is truncated in practical calculations
(only those
operators whose dimension is smaller than some number are
retained). The uncertainty introduced in this way in the Euclidean
calculation is expected to be less than the last term retained.
As has been already mentioned the effect of the high order `tail' of
the power
terms is exponential. The exponential terms are conceptually
connected with  the divergence
of OPE for
high-dimension operators and do not show up in the truncated OPE.
One should proceed, however, with extreme caution estimating the
uncertainty due to the high order power terms in the Minkowski
domain (see below). The existing theory
tells us nothing about when the exponential
terms  become negligibly small. At this point
we have to rely on indirect methods and phenomenological
information existing in this or that channel.

Let us discuss a serious theoretical
question -- whether it is legitimate at all to retain in the expansion of
the correlation functions logarithmic, power and exponential terms
simultaneously. Indeed, formally any $1/(\ln Q)^l$ term is
parametrically larger that, say, $1/Q^4$, and, moreover, any power
term is parametrically larger than exponential ones. If so, the
theoretical uncertainty due to truncation of the logarithmic series --
and we, of course, have to truncate it at some finite order -- is
formally larger than even the lowest-order power correction.

There are two arguments which seemingly justify the procedure of
retaining power terms simultaneously with the truncated
perturbative series. First, in many instances power terms have
distinct origin, and the corresponding effects do not show up in
perturbation theory. This is
valid, for instance, in all cases where the chiral quark condensate is
involved. Another example is due to four-fermion operators
responsible for the Pauli interference in the non-leptonic weak
decays of heavy flavors \cite{VS}. Second, even if the power terms
are not
different in their structure from those occurring in perturbation
theory, there is a strong numeric enhancement of the power
terms inherent to QCD whose origin is not completely clear at the
moment. Thus, estimating the uncertainty due to infrared
renormalons we get $Q^{-4}$ with the coefficient much smaller
than that appearing in Wilsonian OPE.

\subsection{Practical version of OPE}

In analytic QCD it is rather difficult to carry out in practice the
consistent
Wilsonian separation procedure outlined above. Therefore, in
applications,  one usually settles for what is called the `practical
version of
OPE'. In the vast majority of applications the practical version works
very well
numerically
\cite{Shif}; notable exceptions are also known, however \cite{Novi1}.

What is the essence of the practical version of OPE? Let us elucidate
this
question using the  example of the previous section.  Assume that we
want to
calculate the vacuum expectation value of $T$ (to be denoted by
$\Pi$)
limiting ourselves to some finite number of loops. For pedagogical
purposes let
us consider the two-loop graphs of Fig. 1. The corresponding
expression for
$\Pi$ has a generic form
\begin{equation}
\Pi (Q^2) \equiv \langle T\rangle = \alpha_s\int dk^2
\frac{1}{k^2}F(k^2/Q^2)
\label{pertpi}
\end{equation}
where $k$ is the virtual gluon momentum (the Wick rotation to the
Euclidean
space is implied) and $F(k^2/Q^2)$ is a dimensionless function
describing the
fermion lines in the graphs of Fig.  1 and emerging after integration
over the
angle variables of the vector $k$. Of course, at $k^2 <\mu_0^2$ the
genuine
expression for $\Pi$ has nothing to do with the integral (\ref{pertpi})
simply
because the gluon Green function has nothing to do with $1/k^2$.
Therefore,
in the genuine OPE one should cut off the integral from below at
$\mu_0^2$
excluding the domain $k^2<\mu_0^2$ from the perturbative
calculation.
This domain must be referred to the vacuum matrix elements of
different
gluon condensates which certainly can not be calculated
perturbatively.
These gluon operators are normalized at $\mu_0^2$, by construction,
and their
coefficients are given, to the leading order in $\alpha_s$, by
the $k^2$ derivatives of the function $F(k^2/Q^2)$ at $k^2=0$. The
Taylor
expansion of this function starts from $k^2$ \cite{Shif}, generating
the
coefficient of $G_{\mu\nu}^2$; the next term of the Taylor expansion
of
$F(k^2/Q^2)$
will give rise to  the coefficient of $(D_\alpha G_{\mu\nu})^2$, and so
on.
Then the coefficient function $C_I$ is given by the integral
(\ref{pertpi}) with
$k^2>\mu_0^2$.

In the practical version of OPE, instead, one does the following.
Let us add  the integral (\ref{pertpi}) in the domain
$0<k^2<\mu_0^2$ to the coefficient $C_I$. One should not think that
this added
integral correctly represents the low $k^2$ part of the graphs at
hand. The
idea is just to add this integral, by definition, and to subtract it. Then
the
coefficient $C_I$, to order ${\cal O}(\alpha_s)$,  is given by full
perturbative graphs of Fig. 1. To avoid double
counting, however, we subtract
the same contribution from the condensates. More exactly,
we subtract {\em almost} the same contribution. Since the sum over
condensates
is truncated at some finite order $n_0$, the best we can do is to
subtract from
the condensate part
the integral
\begin{equation}
\alpha_s\int^{\mu_0^2}dk^2 \frac{1}{k^2}\sum_{n=2}^{n_0}
\frac{1}{n!} F^{(n)}(k^2/Q^2) .
\label{subtr}
\end{equation}
In this way  we subtract from each condensate its `one-loop
perturbative
value'.

It is quite clear that this strategy of converting the genuine Wilson
OPE
into a practical version (i) leads to a loss of factorization of short and
large
distance contributions inherent to Wilson's OPE; (ii) can not be
systematically
generalized to any number of loops since there is no way to
unambiguously
define the integrand to all orders in the small $k^2$ domain (iii) only
approximately avoids
double counting  due to the necessity of truncating the condensate
series at a
finite order (the lower the order we truncate the larger error is
introduced;
likewise, the larger value of $\mu_0^2$ is chosen the larger error is
introduced).

Therefore, the practical version described above is useful (i.e.
sufficiently
accurate numerically)  only provided  $\mu_0^2$ can be chosen
small enough
to ensure that the `one-loop perturbative' contributions to the
condensates are
much smaller than their genuine values and, at the same time,
$\alpha_s
(\mu_0^2)/\pi$ is small enough for the expansion to make sense.
The existence of such a window is not granted {\em apriori} and is a
very fortunate feature of QCD. The practical version obviously
ignores the semihard contributions to the coefficient functions
discussed above. Working within this version we automatically
limit ourselves to a finite order in $\alpha_s$ forbidding any
questions concerning any aspects of the high order behavior.

\subsection{Consequences of OPE in the Minkowski domain}

Now, let us recall that in the problems at hand we are not interested
in the
correlation
functions at the Euclidean values of $q^2$ {\em per se}.  The
physically observable quantity is the imaginary part. The question is
what can be predicted about the imaginary part from the Euclidean
expansion of the type discussed above. In other
words, we would like to reinterpret the expansion in terms
of the imaginary parts it corresponds to.

If in the Euclidean expansion the notion of the soft and hard lines
means small and large off-shellness, respectively,  one can not
extend this classification to the Minkowski domain. This is especially
clear for the perturbative contribution which, by means of the
Cutkosky rule, is obtained by putting all quarks and gluons in the
intermediate state on shell.

Of course, quarks and gluons are confined and have no mass shell,
literally speaking. The genuine imaginary part is saturated by
hadrons, not quarks and gluons. At large $s$ (where $s=q^2$)
the numerical value of the genuine
imaginary part coincides with that obtained by calculating with
quarks and
gluons only approximately. How large is the actual deviation between
the genuine imaginary part and that inferred from the quark-gluon
calculation in the Euclidean domain? This is the key question the QCD
practitioners have to address sooner or later. I will suggest some
tentative answers.

Let us first explain what is meant when one speaks about
the quark-gluon imaginary part. The starting point is the Euclidean
calculation. Consider first the perturbative series for $\Pi (Q^2)$
truncated at some finite order below the critical order where
the perturbation theory becomes senseless. The successive terms in
this series have the form
$$
\frac{1}{\ln Q^2}, \,\,\, \frac{1}{(\ln Q^2)^2}, ...,\,\,\, \mbox{and}\,\,\,
\frac{\ln \ln Q^2}{(\ln Q^2)^2}\,\,\, \mbox{and so on}.
$$
Now, in each separate term of this series we substitute
$Q^2\rightarrow - s$ and then take the imaginary part.

We repeat the procedure with the condensate terms, again
truncated at some finite order below the point where the asymptotic
divergence shows up.  In the Euclidean domain  a
generic contribution due to higher dimensional operators is
$$
\frac{1}{(Q^2)^k} \times [(\ln Q^2)^\gamma + ...]
$$
where $k$ is an integer number related to the (canonical) dimension
of the operator at hand.
  It translates into
$$
\frac{\gamma\pi}{(s)^k}\times [(\ln s)^{\gamma -1} +...]
$$
in the imaginary
part. Notice that the imaginary parts concentrated at $s=0$ (i.e.
$\delta(s)$ and its derivatives) do not manifest themselves at large
$s$ in the truncated series.  Notice also that  in the absence of the
logarithmic factors (i.e.
for  $\gamma =0$) we get no imaginary part at large
$s$. This feature is specific to the example at hand and does not
extend, generally speaking, to other problems.

Summarizing, we do a straightforward
analytical continuation, term by term. The sum of the imaginary
parts obtained in this way will be referred to as the {\em
quark-gluon}
cross section. This quantity will serve as a reference quantity in
formulating the duality relations. When one says that the hadron
cross section is dual to the quark-gluon one the latter must be
calculated by virtue of the procedure described above.

Defining the analytic continuation to the Minkowski domain in this
way it
is not difficult to see that those lines which were far
off shell in the Euclidean calculation remain hard in the sense that
now they are either still far off shell or on shell but carry large
components of the four-momenta, scaling like $\sqrt s$.

The contributions left aside in the above procedure are related, at
least at a conceptual level, to high-order tails of the both series, the
logarithmic
$\alpha_s$ series  and the power (condensate) one.
These tails generate negligibly small terms in the Euclidean domain,
falling off at large $Q^2$ as a high (and, generically, non-integer)
power of $1/Q^2$ or exponentially,
$\sim \exp (-Q\rho_0)$.

Both can be visualized through instantons. In the first case we deal
with the direct instanton contribution integrated over $\rho$ (the
$\rho$ integration is saturated at $\rho\sim Q^{-1}$). In the second
case $\rho$ is to be considered as a fixed parameter not scaling with
$Q^{-1}$. As we will see below, the fixed-size instanton is actually
relevant in both cases, so we start our discussion assuming
that $\rho$ is fixed. The corresponding contribution to
$\Pi (Q^2)$ was found long ago \cite{AG,DS}. In the Euclidean domain
it is representable as an integral over an auxiliary parameter over a
McDonald function. We do not intend here to rely on specific details
of the instanton expression; rather, we would like to abstract
general features inherent to mechanisms of this type. To this end we
will follow the suggestion of Ref. \cite{DS2}. The following
observation was made in this work: if the condensate terms
correspond, in the coordinate space, to  singularities of
$\Pi (x)$ at $x=0$, the instanton-type contributions are connected
with
the singularities of $\Pi (x)$ located at a finite distance from the
origin; the distance from the origin plays the same role as the
instanton radius. For our purposes it is reasonable to assume that
these singularities have the simplest possible structure, namely,
$$
\frac{1}{x^2+\rho^2} ,\,\,\, \ln{(x^2+\rho^2)},\,\,\,
({x^2+\rho^2}) \ln{(x^2+\rho^2)},\,\,\, \mbox{and so on}.
$$
The Fourier transforms of these expressions have the generic form
$$
(Q\rho)^{-n}K_n(Q\rho),\,\,\, n= 1,2,...
$$
where $K_n$ is the McDonald function.

As was expected, at large Euclidean $Q^2$ the corresponding
contribution dies off exponentially,
\begin{equation}
\Delta\Pi = (Q\rho)^{-n}K_n(Q\rho)\propto (Q\rho)^{-n-1/2}{\rm e}^{-
Q\rho} \, ,
\label{deltapi}
\end{equation}
in full accord with intuition regarding transmitting a large
momentum through a soft field fluctuation. We are interested,
however, in the Minkowski values of the momentum. The question is
what is the impact of the contribution above on Im$\Pi$.

Analytically continuing $\Delta\Pi $ given in Eq. (\ref{deltapi})
to the Minkowski space ($Q\rightarrow iQ$) we arrive at
\begin{equation}
\mbox{Im}\, \Delta\Pi  = (-1)^{n+1}\frac{\pi}{2}
(\sqrt{s}\rho )^{-n} J_n (\sqrt{s}\rho )
\propto
(-1)^{n+1}\frac{\pi}{2}(\sqrt{s}\rho )^{-n-1/2}
\cos (\sqrt{s}\rho - \delta_n )\, ,
\label{imdelta}
\end{equation}
where $J_n$ is the Bessel function. We see that the imaginary
part falls off rather slowly, as a modest power of $1/\sqrt{s}$,
and oscillates.

This result may or may not seem surprising. It tells us
that
with no further dynamical input, on general grounds alone, one can
not rule out large  violations of duality. If this were the end of the
story the operator product expansion would be, to a large extent,
useless in all problems where one needs to predict quantities
referring to the Minkowski kinematics, such as decay probabilities,
spectra, etc.

Fortunately, the situation is not as bad as it might seem at first sight.
Indeed,
so far we were discussing transmitting a large momentum through
a fixed-size fluctuation. In QCD the size of the fluctuation is not
fixed; rather, we integrate it over with a weight function
which depends on $\rho$.

The result of this integration depends on the choice of the weight
function $w(\rho )$. Leaving aside inessential details we are
basically left with
two distinct possibilities -- one will represent the tail of the
$\alpha_s$ series, the other the tail of  the
condensate series. The first option is related to the
small-size direct instantons $\rho\sim 1/\sqrt{s} $; it generates
a duality violating contribution falling off as a high power of
$1/\sqrt{s}$. The second type of the duality-violating contribution,
falling off exponentially with $\sqrt{s}$, is due to instantons
whose size is smeared around some $\rho_0$ (i.e. parametrically
small
$\rho$ are excluded).

Let us start from the small-size instantons. If $\sqrt{s}$ is
parametrically large and $\rho\sim 1/\sqrt{s} $ is parametrically
small one can use the dilute instanton gas approximation
\cite{CDG}.  Then the instanton density $d(\rho )$ (related to the
weight function mentioned above) has the form
\begin{equation}
d(\rho ) = (\Lambda_{\rm QCD}\rho )^b
\end{equation}
where $b$ is the first coefficient in the Gell-Mann-Low function.
For three (almost) massless quarks $b=9$.
For small $\rho$ the weight function $w(\rho )$ has the same form
as $d(\rho )$,
\begin{equation}
w(\rho ) = (\Lambda_{\rm QCD}\rho )^\gamma f(\rho ),
\label{weight}
\end{equation}
but the exponent $\gamma$ depends
on dynamical details. Say, if we treat three light quarks as
 massless (i.e. neglect their mechanical masses) then
$w(\rho )$ is proportional to the cube of the quark condensate
\cite{InSVZ}  and $\gamma = 18$. If the strange quark mass is
considered as a parameter  $\lsim\Lambda_{\rm QCD}$ but
the mechanical masses of $u$ and $d$ are still neglected
$w(\rho )$ is proportional to the square of the quark condensate and
$m_s$, and
and $\gamma = 16$. If all three light quarks are treated as massive
$\gamma = 12$. Even following the latter (quite unrealistic)
treatment we can safely say that $\gamma$ is large numerically.
Let us remind that generically $\gamma$ need not be integer. For
instance,  if one considers very small values of $\rho$ where the
$c$
quark can be treated as light (i.e. $\rho m_c \ll1$) one should use
$\gamma = 55/3$ in the chiral limit with respect to $u$, $d$, $s$
quarks.  This example is quite academic, though, since at such small
values of $\rho$ the weight function is so  strongly
suppressed that the corresponding contribution can hardly be
observable in hadronic processes in the foreseeable future.

Since our purpose  is mostly illustrative we will not restrict
ourselves to a specific value of $\gamma$; rather it will be treated
as a generic numerical parameter lying somewhere in between 10
and 20.

The function $f(\rho )$ in Eq. (\ref{weight}) is a cut off function
suppressing large values of $\rho$. It is equal to unity
in the dilute instanton gas approximation. In fact, one does not need
any cut off if calculations are done at large Euclidean momenta $Q^2$.
The integral is automatically convergent at $\rho \sim 1/Q$ (see
Eq. (\ref{deltapi})).
Since we deal now with the imaginary parts that fall off at large
$\sqrt{s}$ only slowly (see Eq. (\ref{imdelta})), to be on the safe side,
we introduce a cut off function. The final result will be essentially
independent
of the choice of $f(\rho)$ provided the cut off is soft enough.
For instance, $f(\rho) =\exp(-\rho /\rho_0)$ is quite suitable.

With this information in hands we proceed to smearing
Im$\Delta\Pi$ given in Eq. (\ref{imdelta}),
\begin{equation}
\langle\mbox{Im}\Delta\Pi\rangle_s
\propto \int (\sqrt{s}\rho )^{-n} J_n (\sqrt{s}\rho ) w(\rho )
\frac{d\rho}{\rho }
\label{integral}
\end{equation}
where the angle brackets denote smearing and the subscript $s$
indicates the we average over the small size fluctuations. Doing the
integral we arrive at
\begin{equation}
\langle\mbox{Im}\Delta\Pi\rangle_s
\propto
\left( \sin\frac{\pi\gamma}{2}\right)  2^{\gamma -n} \Gamma
(\frac{\gamma}{2})  \Gamma (\frac{\gamma -2n}{2})
\times\left(
\frac{\Lambda_{\rm QCD}}{\sqrt{s}}\right)^\gamma
\label{1/s}
\end{equation}
plus subleading in $1/\sqrt{s}$ terms.  This expression assumes that
$\sqrt{s}\rho_0\gg\gamma$, and characteristic values of $\rho$
saturating the integral (\ref{integral})
are of order
$\rho\sim\gamma /\sqrt{s}$. The very same expression for
the imaginary part could be obtained by integrating the Euclidean
$\Delta\Pi$ over $\rho$
with the subsequent analytic continuation to the Minkowski domain
and separation of the imaginary part.

What lessons can we learn from this simple exercise that describes,
at a qualitative level, transmitting large momenta through a
small-size fluctuation?  First, this particular duality-violating
contribution to the imaginary part falls of as a  high-power of
$1/\sqrt{s}$ at
large $s$.  In order to be able to single out the small-size
fluctuations one must choose $\sqrt{s} > \gamma /\rho_0$.  Even if
$\rho_0^{-1}$ is set at 0.5 GeV (in fact, this parameter is probably
larger \cite{shuryak}) the small-size direct instantons can show up
at energies $\sqrt{s} > $ 5 GeV.  At these energies
\begin{equation}
\langle\mbox{Im}\Delta\Pi\rangle_s
\sim \left( \frac{\gamma\Lambda_{\rm
QCD}}{\sqrt{s}}\right)^\gamma \lsim (1/2)^\gamma
\label{numim}
\end{equation}
and this effect is unimportant for all practical purposes. Parameter
$\Lambda_{\rm QCD}$ in Eq. (\ref{numim}) can be hardly pinned
down better than up to a factor of two at present but this is
quite irrelevant -- the nature of the $s$ dependence in
 Eq. (\ref{numim}) is such that we can safely approximate it by the
step function. If $\sqrt{s}$ is at least factor of two higher
than $\gamma\Lambda_{\rm QCD}$ the result is practically zero;
shifting slightly towards lower values of $\sqrt{s}$ we formally
observe
an almost immediate explosion in Eq. (\ref{numim}). This explosion
happens, however, outside the domain where one can speak about
the small-size fluctuations. The conclusion is as follows: immediately
above a critical point $\langle\mbox{Im}\Delta\Pi\rangle_s$
practically vanishes.  Below the critical point no reliable predictions
exist in the Minkowski domain.

Let us discuss now another contribution which is formally much
smaller
than that of the small-size instantons, namely the contribution
coming
from the vacuum medium where the size of the fluctuations can not
be
arbitrarily small. This is presumably the dominant component of the
vacuum fields  responsible for the most salient features
of the hadronic processes. Being formally exponentially suppressed
this
mechanism at the moment seems to be the leading source of the
duality
violations in all processes of practical interest.
To have something particular in mind one can
think of the instanton liquid model \cite{shuryak}, although this
model is certainly
not
singled out  in the context under discussion. The general features will
be
inherent to any mechanism of this type.

As was already demonstrated, the contribution of the fixed-size
instanton
falls off very slowly
in the Minkowski domain. Certainly, it is not realistic to saturate by
fixed-size
fluctuations;  the ensemble of fluctuations in the genuine QCD
vacuum
is characterized by a distribution in the sizes centered near some
typical
size $\rho_0$. Let us assume that this distribution falls off steeply
at smaller and larger values of $\rho$, with a width $\Delta$. A
typical
distribution of this type can be parametrized as follows:
\begin{equation}
w(\rho ) = {\cal N} \exp\{-\frac{\alpha}{\rho}-\beta\rho\}
 \label{fsid}
\end{equation}
where ${\cal N}$ is a normalization constant,
\begin{equation}
\alpha=\frac{\rho_0^3}{\Delta^2},\,\,\,\, \beta=
\frac{\rho_0}{\Delta^2} ,
\label{constants}
\end{equation}
$\rho_0$ is the center of the distribution and $\Delta$ is its width.

Averaging  $\mbox{Im}\, \Delta\Pi $ over this weight function
one obtains (for $n=1$)
\begin{equation}
\langle\mbox{Im}\, \Delta\Pi \rangle_l
={\cal N} 2E^{-1}J_1
\{ \sqrt{2\alpha}[\sqrt{\beta^2+E^2}-\beta\}^{1/2}]
K_1\{\sqrt{2\alpha}[\sqrt{\beta^2+E^2}+\beta\}^{1/2}]
\label{usrosc}
\end{equation}
(similar expressions for $n>1$ can be derived by differentiating with
respect to $\alpha$; the subscript $l$ indicates that we average over
the large size fluctuations).  If
\begin{equation}
E\gg \frac{1}{\Delta}\frac{\rho_0}{\Delta}
\label{egg}
\end{equation}
the smeared imaginary part reduces to
\begin{equation}
\langle\mbox{Im}\, \Delta\Pi \rangle_l \rightarrow
2{\cal N} E^{-1} J_1 (\sqrt{2E\rho_0}\frac{\rho_0}{\Delta} )K_1
(\sqrt{2E\rho_0}\frac{\rho_0}{\Delta} )\, .
\label{asus}
\end{equation}

It is quite clear that the exponential suppression
of $\langle\mbox{Im}\, \Delta\Pi \rangle_l $ that emerged after
the smearing is due to the oscillations of the original imaginary part
generated by the fixed-size fluctuation. If the weight function is
narrow ($\Delta\ll\rho_0$) this suppression starts at high energies,
see Eq. (\ref{egg}). In the limit when $\Delta\rightarrow 0$ with $E$
fixed the exponential suppression disappears from Eq. (\ref{usrosc}),
and we return back to the original oscillating imaginary part. Notice
also that the exponent at $E\gg \frac{1}{\Delta}\frac{\rho_0}{\Delta}
$ is different
from that one deals with in Euclidean domain ($\sqrt{E}$ {\em
versus} $Q$). Moreover, apart from the exponential suppression we
observe residual oscillations as well. Although the particular
results (\ref{egg}), (\ref{asus}) for the duality violating contributions
were obtained
by using instanton-inspired calculations (and are dependent on the
choice of the weight function) the general features of the
results --  an exponential suppression  starting at a point
correlated with the width of the smearing function plus residual
oscillations -- are presumably common to all mechanisms and will be
confirmed in the future solution of QCD (if it is found) and/or
phenomenologically. I make a  conjecture that the exponentially
suppressed duality violating contributions have a generic form
\begin{equation}
\exp(-f(E)), \,\,\,  f(E)\rightarrow kE^\sigma \,\,\, \mbox{at large}\,
\,\, E
\label{expo}
\end{equation}
where the critical index $\sigma < 1$.
Shortly  we will discover the same
qualitative picture
in a totally different context.

It is worth emphasizing once more that although at academically
high energies the exponentially suppressed contribution dies off
faster than that of the small-size instantons in all practical problems
the latter is totally negligible once we cross the critical point while
the former, being a less steeper function, can play a role.

\subsection{Further explorations of deviations from duality}

Let us consider now a model spectral density suggested in
Ref.  \cite{ShifCAQCD}.  Although this model  is not derived as a
solution of
QCD it satisfies all general properties of QCD we are aware of today
and seems to be very instructive in studying the duality violations.
It will be seen that qualitatively it nicely matches with the
discussion above.

To begin with we remind the model and its motivations.
The model can be formulated, in a most straightforward way, for the
spectral density associated with one heavy and one light quark in the
limit when the heavy quark mass tends to infinity. Simplifications
occur due to the fact that the heavy quark Green function then is
trivially known. Specifically,  let us consider  heavy-to-light
quark currents
$$
J_S=\bar Q q,\,\,\, J_P=\bar Q\gamma_5 q\, ;
$$
\beq
J_1 = \frac{1}{2}(J_S+J_P)=\bar Q\frac{1}{2}(1+\gamma_5 )q,\,\,\,
J_2 = \frac{1}{2}(J_S^\dagger-J_P^\dagger)=
\bar q\frac{1}{2}(1+\gamma_5 )Q \, .
\eeq
 assuming that $m_Q\rightarrow\infty$ and $m_q\rightarrow 0$.
Note that $J_2$ is {\em not} hermitean conjugate to
$J_1$; rather  $J_2$ is a chiral partner to $J_1^\dagger$. Therefore,
the correlation function
\beq
\Pi = i\int {\rm e}^{ikx} dx \langle vac |T\{J_1(x), J_2(0)\}|vac\rangle
\eeq
vanishes in perturbation theory if $m_q$ is put  to zero.

First of all, we
should choose the external momentum $k$ in a most advantageous
way.  If
$m_Q\rightarrow\infty$ it is clear that the optimal reference point
lies
``slightly" below threshold,
$$
k_0 = m_Q -\epsilon,  \,\,\, \vec k =0,
$$
where $\epsilon ={\cal O}(\Lambda_{\rm QCD})$.

Taking  the limit $m_Q\rightarrow\infty$ and neglecting the hard
gluon exchanges one can rewrite the correlation function $\Pi
(\epsilon )$ at positive (Euclidean) values of $\epsilon$ in the
following way \cite{ShifCAQCD}:
\begin{equation}
\Pi (\epsilon ) =\frac{1}{4} \int_0^\infty e^{-\epsilon\tau}d\tau
\langle\bar q (\tau) \exp\{\int_0^\tau  igA_0 (t ) dt\} q(0
)\rangle
\label{WL}
\end{equation}
where the angle brackets denote the vacuum averaging.  The
correlation function of interest is thus nothing else than the Laplace
transform
of the Wilson
line (in the time direction) with the light quark fields at the end
points.

If $\epsilon\gg\Lambda_{\rm QCD}$ one can expand
Eq. (\ref{WL}) in $1/\epsilon$,
\beq
\Pi (\epsilon )
= \frac{1}{4\epsilon}
\left[
\langle \bar q q \rangle -\frac{1}{\epsilon^2}
\langle \bar q {\cal P}_0^2q \rangle +
\frac{1}{\epsilon^4}
\langle \bar q {\cal P}_0^4 q \rangle -
\frac{1}{\epsilon^6}
\langle \bar q {\cal P}_0^6 q \rangle  + ...\right] ,
\label{WLE}
\eeq
where ${\cal P}_0$ is the time component of the ({\em Euclidean})
momentum
operator.

{}From Eq. (\ref{WLE}) it is perfectly clear that the operator product
expansion
for $\Pi (\epsilon )$ must be asymptotic; otherwise it would define
an odd
function of $\epsilon$, which is obvious nonsense.
Indeed, at positive $\epsilon$ we are below the cut, and $ \Pi
(\epsilon )$
is analytic in $\epsilon$. At negative $\epsilon$, however, we sit
right on the cut generated by the intermediate physical states
produced by the currents $J_{1,2}$, and  the correlation function
$\Pi (\epsilon )$ develops an imaginary part, a discontinuity across
the cut. Qualitatively we have a pretty good idea of how
${\rm Im}\Pi$ looks like. We will return to this issue later on and
now
observe only that if the series in (\ref{WLE}) is truncated at any
finite
order
the only imaginary part we obtain from Eq. (\ref{WLE}) is
concentrated at
$\epsilon = 0$.
In accordance with our formulation of duality we conclude
that at large positive  values of $E$ ($E\equiv -\epsilon$),
$
E\gg \Lambda_{\rm QCD} ,
$
the physical spectral density is predicted to {\em  vanish}.

While the exact spectral density is unknown, of course, a model
which should be close to it, at least qualitatively, has been
suggested in Ref. \cite{ShifCAQCD}.  On general grounds we know
that at large $\tau$
the Wilson line $\langle\bar q (\tau) \exp\{\int_0^\tau  igA_0 (t )
dt\} q(0
)\rangle$ must be proportional\footnote{Warning: in many
publications it is
erroneously assumed that $\langle\bar q (\tau) \exp\{\int_0^\tau
igA_0 (t )
dt\} q(0)\rangle \propto\exp(-C\tau^2)$ which is incompatible with
general
principles.}
 to $\exp\{ -\bar\Lambda\tau\}$; moreover, this function should
have no
singularities in the complex plane except on the imaginary axis. At
small
$\tau$ it must be expandable in powers of $\tau$ and this expansion
must
run over only even powers of $\tau$ (this expansion generates Eq.
(\ref{WLE})).  The simplest choice satisfying the above criteria is
\begin{equation}
\langle\bar q (\tau) \exp\{\int_0^\tau  igA_0 (t ) dt\} q(0
)\rangle =\langle\bar q q\rangle\times
\frac{1}{\cosh\bar\Lambda\tau} .
\label{cosh}
\end{equation}

This expression for the Wilson line generates a correlation function
$\Pi (\epsilon )$,
\begin{equation}
\Pi (\epsilon ) =\frac{1}{4\bar\Lambda}\langle\bar q q\rangle \beta
\left(\frac{\epsilon +\bar\Lambda}{2\bar\Lambda}\right)
\label{beta}
\end{equation}
where $\beta$ is related to Euler's $\psi$ function,
$$
\beta (x) =\frac{1}{2}\left[ \psi\left(\frac{x+1}{2}\right)
- \psi\left(\frac{x}{2}\right)\right] =\sum_{k=0}^\infty\frac{(-
1)^k}{x+k}\,\,\,  .
$$

We pause here to make a digression illustrating the degree to which
realistic details can be expected to be reproduced by the model
above. Accepting Eq. (\ref{cosh})
we made a specific assumption about the behavior of the matrix
elements
appearing in Eq. (\ref{WLE}) -- as dimension of the operators grows
their
matrix elements grow factorially,
\beq
\langle \bar q({\cal P}_0)^{2n} q \rangle\sim
\langle \bar q q \rangle (\Lambda_{\rm QCD}^2)^n C^{2n}(2n)!\,  ,
\eeq
where $C$ is a numerical constant.  Although very little is known for
fact for high-dimension operators one can confront the dimension-5
operator with information that already exists. Indeed,
\begin{equation}
\langle\bar q {\cal P}_0^2 q\rangle =
\frac{1}{8} \langle\bar q ig G_{\mu\nu}^a\sigma_{\mu\nu}t^a
q\rangle = \frac{1}{8}m_0^2\langle\bar q  q\rangle
\label{Iof}
\end{equation}
where $m_0^2$ is a parameter introduced in \cite{Ioffe}
\footnote{For a review see the Reprint Volume cited in Ref.
\cite{Shif}; note that Eq. (2.4) on page 22 of the above Volume
contains a misprint, $-g$ must be substituted by $ig$; moreover,
$\sigma_{\mu\nu}$ is understood as
$(1/2)[\gamma_\mu\gamma_\nu ]$.},  and we accounted for the fact
that ${\cal P}_0$ is defined in Euclidean. By comparing the
first subleading term in the large $\epsilon$ expansion of
Eq. (\ref{beta}) with the general expansion (\ref{WLE}) we
establish that (i) the sign is correct; (ii) ${\bar\Lambda}^2=
m_0^2/8$. Phenomenologically $m_0^2$
is close to $0.8$ GeV$^2$ \cite{Ioffe}, and  we conclude that in our
model
$$
\bar\Lambda\approx 0.31 \mbox{GeV} .
$$
This number is in the right ballpark; perhaps, a factor of $\sim$ 1.5
smaller
than the expected phenomenological value of $\bar\Lambda$.

The correlation function (\ref{beta}) we end up with is a sum of
simple poles
at
$$
E\equiv - \epsilon = 1,  3,  5,  7, ....
$$
(in the remainder of this section we put $\bar\Lambda = 1$; thus all
energies
will be  measured in these units), all residues are equal in the
absolute value
and are sign-alternating. We remind that Im$\Pi (E)$  is actually the
difference between
the spectral densities in the scalar and pseudoscalar channels
and, therefore, need not be positive.

The above model perfectly matches a picture one expects to get in
 the multicolor QCD, with $N_c\rightarrow\infty$:
two combs of infinitely narrow peaks sitting, back-to back,
on top of each other.  Let us defer for a while
the discussion of how the spectral density evolves from $N_c=\infty$
to $N_c= 3$ and study now the implications of this picture for
duality.

The spectral density at $E \gg\Lambda_{\rm QCD}$
stemming
from Eq.
(\ref{beta})
\begin{equation}
\mbox{Im}\Pi =\frac{\pi}{2}\langle\bar q q\rangle
\sum_{k=0}^\infty (-1)^k\delta (E- (1+2k))\,\, ,
\end{equation}
taken at its face
value, does not vanish. We understand, however, that under the
circumstances duality should be applied to the average spectral
density
rather than to Im$\Pi (E)$ itself.  If we merely average Im$\Pi (E)$
in the interval from $E_1$ to $E_2$ we get a sign-alternating result
of order $1/(E_2-E_1)$ -- i.e. a huge deviation from duality which
does not even die off as $E\rightarrow\infty$ as long as the size of
the smearing interval is kept fixed!

If this were the actual situation in QCD any OPE-based expansions in
the Minkowski domain would be hopeless.
For instance, why then bother about calculating the $1/m_Q^2$
corrections to the lepton spectrum  in $B\rightarrow l\nu X_u$
due to $\mu_G^2$ and $\mu_\pi^2$ \cite{A4} if uncontrollable
deviations may
die off slower than $1/m_Q^2$?
Fortunately, at large
energies QCD with $N_c=3$ does not look at all like its multicolor
limit; the limits $E\rightarrow\infty$ and $N_c\rightarrow\infty$
are not commutative.
If the number of colors decreases from  infinity to 3  the spectral
density Im$\Pi (E)$ at large $E$ experiences a dramatic evolution.
A  comb of
delta functions is converted into a smooth function. The details of the
smearing mechanism are foggy but qualitatively one may conjecture
the following. At $N_c=\infty$ the correlation functions in QCD are
saturated by a masterfield \cite{Witten79}, a classical field
configuration which does not fluctuate. This freezes the scale and
makes the calculation in a sense similar to that with the fixed-size
instanton.  With $N_c$ decreasing, the field configurations saturating
the functional integral start fluctuating; correspondingly,  a
distribution of scales emerges.  The correlation function
corresponding to $N_c=3$ can be obtained, qualitatively,  by
substituting $\epsilon \rightarrow\epsilon \rho$ in Eq. (\ref{beta})
and integrating over $\rho$ with a weight function similar to
(\ref{fsid}) centered at $\rho =1$. In numerical exercises below we
will consider  the weight function (\ref{fsid}) as a particular
example, with
$$\alpha=\beta=\frac{1}{\Delta^2},\,\,\,  {\cal N}=\frac{1}{2K_1
(2/\Delta^2)} \, ,
$$
where $\Delta$ is the width of the distribution,
$\Delta ={\cal O}(1/N_c)$.

Thus, our model spectral density becomes
$$
\mbox{Im}\Pi (E) = \int_0^\infty d\rho w(\rho )
\frac{\pi}{2}\langle\bar q q\rangle
\sum_{k=0}^\infty (-1)^k\delta\left( E\rho - (1+2k) \right)
$$
\begin{equation}
=\frac{\pi}{2E}\langle\bar q q\rangle
\sum_{k=0}^\infty (-1)^k w\left(\frac{1+2k}{E}\right) .
\label{sumw}
\end{equation}

Analytically it is not difficult to prove that this imaginary part falls
off faster than any given power of $1/E$.
Indeed, let us focus on the function
\begin{equation}
\sigma (E) =
\sum_{k=0}^\infty (-1)^k w\left(\frac{1+2k}{E}\right) =
\sum_{k=0}^\infty \left[ w\left(\frac{1+4k}{E}\right)
-w\left(\frac{3+4k}{E}\right)\right] \, .
\end{equation}
To show that at large $E$ terms ${\cal O}(1/E)$ are absent in $\sigma
(E)$ we represent
$$
w\left(\frac{3+4k}{E}\right) =\frac{1}{2}\left[
w\left(\frac{1+4k}{E}\right)J+ w\left(\frac{5+4k}{E}\right)\right]
+{\cal O}(1/E^2) .
$$
Then
$$
\sigma (E) =\frac{1}{2}\sum_{k=0}^\infty \left[
w\left(\frac{1+4k}{E}\right)
-w\left(\frac{5+4k}{E}\right)\right] + {\cal O}(1/E^2)
$$
$$
= w(1/E) + {\cal O}(1/E^2) .
$$
Moreover, in the next order one uses
$$
w\left(\frac{3+4k}{E}\right) =\frac{3}{8}\left[
w\left(\frac{1+4k}{E}\right)J+ 2w\left(\frac{5+4k}{E}\right)
-\frac{1}{3}w\left(\frac{9+4k}{E}\right)J\right]
+{\cal O}(1/E^3) .
$$
an so on.

Hence, the suppression is exponential, modulated by oscillations.
These  facts are  quite general and do not depend on the particular
choice of the weight function provided that the weight function falls
off faster than any power of $\rho$ at small $\rho$. For instance, to
prove that an infinite number of oscillations must necessarily take
place  in the imaginary part
we write the dispersion relation for $\Pi (\epsilon )$, expand it in
$1/\epsilon$
and compare the result of this expansion with Eq. (\ref{WLE}).
The absence of all even powers in $1/\epsilon$ requires
an infinite number of oscillations in Im$\Pi (E)$.

I was unable to find analytically the type of the exponential
suppression.  Numerically it seems that Eq. (\ref{expo}) goes through,
with the index $\sigma$ depending on the width $\Delta$. (Thus, at
$\Delta = 1/3$ we get $\sigma \approx 0.7$).   The plots
of Im$\Pi (E)$ for $\Delta =0$, $\Delta = 1/3$ and $\Delta = 1/2$ are
presented on
Fig. 4.  We see that the heights of the oscillations rapidly fall off
starting with some boundary energy which, in turn, depends on
$\Delta$. Duality is restored with the exponential accuracy at
$E\gg 1/\Delta$. Below this boundary value the resonance peaks are
conspicuous. Notice also that  the length of oscillations grows with
energy. At $\Delta\rightarrow 0$ the boundary value of energy
shifts to infinity,
and we return back to the infinite comb of the infinitely narrow
peaks.

Theoretically the precise shape of the fall off of Im$\Pi (E)$
at large $E$ is correlated with the divergence of the high-order terms
in Eq. (\ref{WLE}). This divergence depends, in turn, on the behavior
of the matrix elements $\langle\bar q {\cal P}_0^{2n} q\rangle$
which is not known.  In our comb-like model (see Eq. (\ref{cosh}))
these matrix elements grow as $(2n)!$. After smearing this formula
changes, the result being dependent on the particular choice of the
smearing weight function. The crucial parameter is $n\Delta$.
As long as this parameter is smaller than unity the leading behavior
remains intact, $(2n)!$. For $n\Delta\gg1$ the regime changes
and $(2n)!$ is substituted by $\Delta^{2n}((2n)!)^2$. We remind that
$\Delta \sim 1/N_c$.
This means that
the terms in the operator product expansion (\ref{WLE}) which are
subleading in $1/N_c$ are more singular in $n$, i.e. their
divergence in high orders of OPE must be stronger. This aspect is not
new, though; the very same situation takes place in the ordinary
perturbation theory \cite{koplik}.

\vspace{0.3cm}
I am grateful to J. Bagger and A. Falk for inviting me to give this talk
at this wonderful Symposium.


\vspace{0.5cm}

\begin{thebibliography}{99}

\bibitem{ShifCAQCD}
M. Shifman, {\em Theory of Preasymptotic Effects
in Weak Inclusive Decays}, in Proc. of  the Workshop {\it Continuous
Advances
in QCD}, ed. A. Smilga, [World Scientific, Singapore, 1994], page 249
[hep-ph/9405246].

\bibitem{HQETrev}
N. Isgur and  M.B. Wise,  in {\em B
Decays}, ed.  S. Stone, 2-nd Edition  (World Scientific, Singapore,
1994), page 231;\\
H. Georgi,   in {\em Perspectives in the Standard Model},  Proceedings
of the Theoretical Advanced Study
Institute, Boulder, Colorado,  1991,  ed. R.K. Ellis, C.T. Hill and
J.D. Lykken,
(World Scientific, Singapore, 1992); {\it Ann. Rev. Nucl. Part. Sci.}
{\bf 43} (1994) 209.

\bibitem{alpha}
E. Bagan, P. Ball, B. Foil and P. Gosdzinsky, Preprint CERN-TH/95-25
[hep-ph/9502338].

\bibitem{alpha2}
M. Neubert,   {\it Phys. Lett.} {\bf B341} (1995) 367;\\
M. Luke, M. Savage and M. Wise,  {\it Phys. Lett.}
{\bf B343} (1995) 329; {\bf B345} (1995) 301.

\bibitem{alpha3}
P. Ball, M. Beneke and V. Braun, Preprint
CERN-TH-95-65 [hep-ph/9503492].

\bibitem{Vbc}
M. Shifman, N. Uraltsev and A. Vainshtein,
{\it Phys. Rev.} {\bf D51} (1995) 2217;\\
I. Bigi, M. Shifman, N. Uraltsev and A. Vainshtein, Preprint
TPI-MINN-94-12-T [hep-ph/9405410].

\bibitem{HQET}
E. Eichten and B. Hill, {\it Phys. Lett.} {\bf B234} (1990) 511;\\
H. Georgi, {\it Phys. Lett.} {\bf B240} (1990) 447.

\bibitem{Wils}
K. Wilson, {\it Phys. Rev.} {\bf 179} (1969) 1499;\\
K. Wilson and J. Kogut, {\it Phys. Reports} {\bf 12} (1974) 75.

\bibitem{muelller}
For a recent review see
A. H. Mueller, {\bf in} {\em Proc. Int. Conf. ``QCD -- 20 Years Later"},
Aachen 1992, eds. P. Zerwas and H. Kastrup, (World Scientific,
Singapore, 1993), vol. 1, page 162.

\bibitem{bigi1}
I. Bigi, M. Shifman, N. Uraltsev and A. Vainshtein, {\it
Phys. Rev. } {\bf D50} (1994) 2234.

\bibitem{bb}
M. Beneke and V. Braun, {\it
Nucl. Phys.} {\bf B426} (1994) 301.

\bibitem{bbz}
M. Beneke, V. Braun  and V. Zakharov, {\it
Phys. Rev. Lett.} {\bf 73} (1994) 3058.

\bibitem{manohar}
M. Luke, A. Manohar, and
M. Savage, Preprint UTPT-94-21 [hep-ph/9407407].

\bibitem{BLM}
S.J. Brodsky, G.P. Lepage and  P.B. Mackenzie, {\it Phys. Rev.} {\bf
D28}
(1983) 228;  G.P. Lepage and  P.B. Mackenzie, {\it Phys. Rev.} {\bf
D48}
(1993) 2250.

\bibitem{nb}
M. Beneke and V. Braun, Preprint DESY-94-200 [hep-ph/9411229];\\
M. Neubert, Preprint CERN-TH.7487/94 [hep-ph/9412265]; Preprint
CERN-TH.7524/94 [hep-ph/9502264];\\
P. Ball, M. Beneke and V. Braun, Preprint CERN-TH/95-26
[hep-ph/9502300].

\bibitem{resum}
J. Chay and S-J. Rey,
Preprint SNUTP-9408 [hep-ph/9404214]; Preprint SNUTP-9454
[hep-ph/9406279];\\
A. Falk and A. Kyatkin, Preprint JHU-TIPAC-950004
[hep-ph/9502248].

\bibitem{NSVZ}
V. Novikov, M. Shifman, A. Vainshtein and V. Zakharov,
{\it Nucl. Phys.} {\bf B249} (1985) 445 [reprinted in
{\it Vacuum Structure and QCD Sum Rules}, ed. M. Shifman,
North-Holland, 1992, page 161].

\bibitem{Nussinov}
S. Nussinov and W. Wetzel, {Phys. Rev.} {\bf D36} (1987) 130.

\bibitem{Voloshin}
M. Voloshin and M. Shifman, {Yad. Fiz.} {\bf 47} (1988) 801
[{Sov. J. Nucl. Phys.} {\bf 47} (1988) 511].

\bibitem{Isgu}
N. Isgur and M. Wise, {\it Phys. Lett.} {\bf B232} (1989) 113;
{\it Phys. Lett.} {\bf B237} (1990) 527.

\bibitem{Luke}
M.E. Luke, {Phys. Lett.,} {\bf B252} (1990) 447.

\bibitem{Chay}
J. Chay, H. Georgi and  B. Grinstein, {Phys. Lett.} {\bf B247} (1990)
399.

\bibitem{Koyrakh}
B. Blok, L. Koyrakh, M. Shifman and A. Vainshtein,
{Phys. Rev.} {\bf D49} (1994) 3356.

\bibitem{Bigi2}
I. Bigi, M. Shifman, N. Uraltsev, A. Vainshtein,
{Int. Journ. Mod. Phys.} {\bf A9} (1994) 2467; \\
M. Voloshin, Preprint TPI-MINN-94/18-T [unpublished].

\bibitem{CGP}
F. Close, G. Gounaris and J. Paschalis, {\it Phys. Lett.} {\bf B149}
(1984) 209.

\bibitem{Ball}
P. Ball and V. Braun, {\it Phys. Rev.} {\bf D49} (1994) 2472;
E. Bagan, P. Ball, V. Braun and P. Gosdzinsky, {\it
Phys. Lett. } {\bf B342} (1995) 362.

\bibitem{mane}
M. Neubert, Preprint CERN-TH/95-107 [hep-ph/9505238].
By quoting this paper I do not imply that I agree with each and
every estimate there; some uncertainties are overestimated, to my
mind.

\bibitem{Fal}
A.  Falk and  M. Neubert, {Phys. Rev.} {\bf D 47} (1993) 2965.

\bibitem{Manne}
T. Mannel,  {\it Phys. Rev.} {\bf D50} (1994) 428.

\bibitem{SUV}
M. Shifman, N. Uraltsev and A. Vainshtein, work in progress that will,
hopefully, be completed one day.

\bibitem{Shif}
M. Shifman, A. Vainshtein and V. Zakharov,
{\it Nucl. Phys.} {\bf B147} (1979) 385;
for a recent review see {\it Vacuum Structure and QCD Sum Rules},
ed. M. Shifman, North-Holland, 1992.

\bibitem{Novi1}
V. Novikov, M. Shifman, A. Vainshtein and V. Zakharov, {\it Nucl.
Phys.} {\bf  B174} (1980) 378; {\it Nucl. Phys}. {\bf B191} (1981)
301; {\it Nucl. Phys}.  {\bf B249} (1985) 445.

\bibitem{HOIA}
For a review see {\em Large Order Behaviour of Perturbation
Theory}, eds. J.C. Le Guillou and J.  Zinn-Justin,  North-Holland,
1990.

\bibitem{renorm}
G. 't Hooft, {\bf in} {\em The Whys Of Subnuclear Physics}, Erice
1977,
ed. A. Zichichi (Plenum, New York, 1977), p. 943;\\
B. Lautrup, {\it Phys. Lett.} {\bf 69B} (1977) 109;\\
G. Parisi, {\it Phys. Lett.} {\bf 76B} (1978) 65; {\it Nucl. Phys.}
{\bf B150} (1979) 163;\\
A. Mueller, {\it Nucl. Phys.} {\bf B250} (1985) 327.

\bibitem{VS}
M. Voloshin and M. Shifman, {\it Yad. Fiz.} {\bf 41} (1985) 187
[{\it Sov. Journ.
Nucl. Phys.} {\bf 41} (1985) 120]; {\it ZhETF} {\bf 91} (1986) 1180
[{\it Sov. Phys. -- JETP} {\bf 64} (1986) 698];\\
N. Bili\'{c}, B. Guberina and J. Trampeti\'{c},
{\it Nucl. Phys.} {\bf B248} (1984) 261.

\bibitem{AG}
N. Andrei and D. Gross, {\it Phys. Rev.}  {\bf D18} (1978) 468.

\bibitem{DS}
M.S. Dubovikov and A.V. Smilga, {\it Nucl. Phys.} {\bf B185}
(1981) 109 [reprinted in
{\it Vacuum Structure and QCD Sum Rules}, ed. M. Shifman,
North-Holland, 1992, page 191].

\bibitem{DS2}
M.S. Dubovikov and A.V. Smilga, {\it Yad. Fiz.} {\bf 37} (1983) 984
[{\it Sov. J. Nucl. Phys.} {\bf 37}
(1983) 585].

\bibitem{CDG}
C. Callan, R. Dashen and D. Gross, {\it Phys. Rev. } {\bf D17}  (1978)
2717.

\bibitem{InSVZ}
M. Shifman, A. Vainshtein and V. Zakharov, {\it Nucl. Phys.}
{\bf B163} (1980) 46.

\bibitem{shuryak}
E. Shuryak, {\it The QCD Vacuum, Hadrons and the Superdense
Matter},
(World Scientific, Singapore, 1988).

\bibitem{Ioffe}
V.M. Belyaev and B.L. Ioffe, {\it ZhETF} {\bf 83} (1982)876;
{\bf 84} (1983) 1236 [{\it Sov. Phys. JETP} {\bf 56} (1982) 493;
{\bf 57} (1983) 716].

\bibitem{A4}
I. Bigi, M. Shifman, N. Uraltsev, A. Vainshtein,
{\it Phys. Rev. Lett.} {\bf 71} (1993) 496.

\bibitem{Witten79}
E. Witten, {\it The 1/N Expansion in Atomic and Particle Physics},
in Proc. Cargese Summer Institute ``Recent Developments in Gauge
Theories" ,
Ed.  G. 't Hooft et al.  N.Y., Plenum Press, 1980.

\bibitem{koplik}
J. Koplik, A. Neveu and S. Nussinov, {\it Nucl. Phys.} {\bf B123}
(1977) 109.

\end{thebibliography}
\end{document}